\documentclass[aps,pra,showpacs,twocolumn,nofootinbib]{revtex4}

\usepackage{dcolumn}
\usepackage{amsmath}

\begin{document}

\title{Energy levels and lifetimes of Gd~IV and enhancement of
the electron dipole moment}

\author{V. A. Dzuba}
\email{dzuba@newt.phys.unsw.edu.au}
\affiliation{School of Physics, University of New South Wales, Sydney 2052,
Australia}

\author{W. R. Johnson}
\email{johnson@nd.edu}
\homepage{www.nd.edu/~johnson}
\author{U. I. Safronova}
\email{usafrono@nd.edu}
\affiliation{
Department of Physics, 225 Nieuwland Science Hall\\
University of Notre Dame, Notre Dame, IN 46566}

\author{O. P. Sushkov}
\email{sushkov@newt.phys.unsw.edu.au}
\affiliation{School of Physics, University of New South Wales, Sydney 2052,
Australia}

\date{\today}

\begin{abstract}

We have calculated energy levels and lifetimes of $4f^7$ and $4f^65d$
configurations of Gd~IV using Hartree-Fock and configuration interaction
methods.
This allows us to reduce significantly the uncertainty of the theoretical
determination of the electron electric dipole moment (EDM) enhancement factor
in this ion and, correspondingly, in gadolinium-containing garnets for which
such measurements were recently proposed.
Our new value for the EDM enhancement factor of Gd$^{+3}$ is $-2.2 \pm 0.5$.
Calculations of energy levels and lifetimes for Eu~III are used to control 
the accuracy.

\end{abstract}


\maketitle

\section{introduction}

There have been recent suggestions by \citet{Lam} and 
\citet{Hun}
 for searches of the electron electric dipole moment (EDM) in solid state
experiments with the compounds Gadolinium Gallium
Garnet Gd$_3$Ga$_5$O$_{12}$ and Gadolinium Iron Garnet
Gd$_3$Fe$_5$O$_{12}$.
It is known that an EDM of a system in a stationary quantum state
violates both  time-reversal (T) and  space-reflection (P)
symmetries. This is why searches for
EDMs of elementary particles, atoms, and molecules are very important for
studies of  violations of fundamental symmetries \cite{KL}.
The best limit on the electron EDM comes from the Berkeley
experiment of \citet{Com} with an atomic Thallium beam, 
$d_e < 1.6\times 10^{-27}$e\, cm.
There are ideas on how to improve the sensitivity substantially working with
PbO \cite{DeM} and YbF \cite{Hinds} molecules.
An alternative that can provide a real breakthrough is to use
solids containing uncompensated electron spins.
This idea was already suggested in \citeyear{Sh} by \citet{Sh}.
Application of a strong electric field to electrons bound within a
solid would align the EDMs of the unpaired electrons. This
should lead to a simultaneous alignment of the electron spins; the
magnetic field arising from this alignment could be detected
experimentally. Another possibility is to polarize electrons by the
external magnetic field. This causes alignment of electron EDMs, and
hence induces a voltage across the sample that could be detected.
An experiment of this kind has been performed with
nickel-zinc ferrite \cite{VK}, however, due to experimental
limitations, the result was not very impressive.
Interest in this approach has been renewed recently owing to the suggestions
by \citet{Lam,Hun} to perform similar experiments with Gadolinium Gallium
Garnet and Gadolinium Iron Garnet, employing new experimental techniques.
The estimates of sensitivity presented in \cite{Lam} look
highly promising; an improvement by several orders of magnitude is feasible.

The first calculations of the expected effects have been performed in recent
papers \citet{Buh,Kuen} using a semiempirical approach.
The compounds under consideration contain Gd$^{3+}$ ions, 
see Ref.~\cite{GGG}, that give the most important contributions to
the effect owing to 
their large nuclear charge. Therefore, from the theoretical point of view, the
problem can be split into two closely connected, but still
distinct, parts. The first part is the evaluation of the EDM of a Gd$^{3+}$ ion
induced by an assumed electron EDM, 
and the second part is an account of the combined
electron-lattice dynamics of the solid.
It has been shown by \citet{Buh} that the main contribution to the Gd$^{3+}$
EDM comes from mixing between $4f$ and $5d$ electrons.  This mixing depends
on the energy separation between $4f^7$ and $4f^65d$ configurations.
There are experimental data on the relevant energy intervals \cite{Gde}.
However, Ref.~\cite{Gde} does not contain identification of all
possible levels; therefore, one cannot rely completely on the data.
It has been pointed out by \cite{Buh,Kuen} that accurate calculations
of the energy levels of the $4f^65d$ configuration of Gd$^{3+}$ 
ion are the needed to improve the accuracy of the EDM calculation.

There were recent measurements of lifetimes of some states of the $4f^65d$
configuration of Eu~III \cite{tau}, which has an electronic structure 
similar to Gd~IV. Therefore, calculations of lifetimes in Eu~III provide
a good test of the accuracy of E1-transition amplitudes which determine both 
lifetimes and the EDM enhancement factor.
In the present work we perform calculations of the energy levels and
lifetimes of both Eu~III and Gd~IV.

\section{Calculation of energies}
\label{energies}

We use two different sets of computer codes to do our calculations.
One is our own configuration interaction program and the second is a
code written by \citet{cowan} and freely available
via the Internet \footnote{ftp://aphysics.lanl.gov/pub/cowan}. 
We use both codes to compute energy levels of
Eu~III and Gd~IV. Calculations for the Eu~III are mostly done to control
the accuracy. These two ions have similar electronic structure, however,
much of the reliable experimental data is available for Eu~III while limited
data are available for Gd~IV.
We restrict our study to the lowest odd configuration $4f^7$ and even
configuration $4f^65d$; these are the configurations most relevant to the
atomic EDM.

We will describe our  approach in detail while restricting our
comments on the Cowan code to few general remarks. Since our calculations
are relativistic, we will use the abbreviation RCI (relativistic configuration
interaction) to refer to them.
We start our calculations
using the relativistic Hartree-Fock (HF) method. Calculations of the
self-consistent field are done for an ion in its ground state.
This is an open-shell system with 7 out of 14 electrons in its
outermost $4f$ subshell. Therefore, we apply 50\% weighting to the
contribution of the $4f$ subshell to the HF potential. This weighting
is further reduced to $\sim 46\%$ when the interaction of a $4f$ electron
with other electrons of the same subshell is calculated
($6/13 \approx 0.46$). Note, that our calculations are relativistic and
we apply the same weighting to both $4f_{5/2}$ and $4f_{7/2}$ subshells.
RCI results for the $4f^65d$ configuration are sensitive to how the
$5d$ state is calculated. It is natural to calculate it in the field
of the $4f^6$ subshell, which means that HF potential is modified by
removing a contribution of one $4f$ electron. However, the
$5d$ state obtained in this way is still not good enough to achieve
accurate energy levels. This is probably because the
self-consistent field is calculated for the configuration $4f^7$ 
and not for $4f^65d$. Therefore, we further modify the $5d$
state by introducing a correction to the HF potential in which this
state is calculated:
\begin{equation}
        \delta V = - \frac{\alpha}{2(a^4+r^4)}.
\label{deltaV}
\end{equation}
Here $\alpha$ is polarizability of an ion in the $4f^6$ configuration,
$a$ is a cut-off parameter introduced to remove the singularity in
the origin. Potential (\ref{deltaV}) describes the effect of core
polarization by the field of external electron. We treat $\alpha$ as
a fitting parameter. Its value ($\alpha = 0.5a_B^3$) has been chosen
to obtain accurate energy levels for Eu~III. The value of $a$ is not very
important because the $5d$ wave function is small at short distances. We use
$a=a_B$. We use the same values of $a$ and $\alpha$ for both Eu~III and Gd~IV.

We now have four single-electron basis states, $4f_{5/2}, 4f_{7/2},
5d_{3/2}$ and $5d_{5/2}$. Many-electron basis states for the RCI
calculations are constructed by distributing seven electrons over
these states in all possible ways. Then, many-electron states of
definite parity and total angular momentum $J$ are constructed. The actual
matrix size depends on the configuration considered and the value of
the total angular momentum $J$; it varies between 1 ($4f^7, J=25/2$) and
377 ($4f^65d, J=9/2$).

Energy intervals in the RCI calculations are sensitive to the value of
Slater integrals ($F_2(4f,4f), F_2(4f,5d)$, etc.).
In the HF approximation, the value of these integrals, and consequently
the energy intervals, are overestimated. This is because of
screening of the Coulomb interaction between valence electrons by core
electrons (see, e.g. \cite{kozlov}). In the present work we include this
screening semi-empirically by introducing screening factor $f_2=0.8$.
The value of this factor was chosen to fit energy intervals in Eu~III.
Thus, in the end, we have two fitting parameters, a core polarizability
$\alpha$ and a screening factor $f_2$. The values of both of these factors
are chosen for Eu~III and then the same values are used for Gd~IV.

Calculations with the Cowan code are very similar to the RCI calculations.
This is also a configuration interaction method, although in its
non-relativistic  realization. There are also two fitting procedures in 
the Cowan code.
One is scaling of the Coulomb integrals by a factor of 0.85. This is very
similar to our screening of Coulomb interaction. Another fitting which we
use in the Cowan code is changing ``by hand'' the average energy of the
ground state configuration. This is equivalent to shifting  all energy
intervals between the ground and excited configuration by the same value.
A few more details about calculations with the Cowan code will be given in
Section \ref{results}.

\section{calculation of lifetimes}

Lifetimes of the three $4f^65d~^8P_{5/2,7/2,9/2}$ states of Eu~III
have been recently measured by \citet{tau}.
Calculations using the Cowan code give values which are about three times
smaller (see Table \ref{lifetime}). It is important to investigate
the source of this discrepancy.

In a single-configuration approximation which we use in the present
work, lifetimes of all states of the $4f^65d$ configuration
are determined by single radial integral
\begin{equation}
        R_{4f,5d} = \int_0^{\infty}R_{4f}(r)R_{5d}(r)r^3dr,
\label{radint}
\end{equation}
and can be presented in a form

\begin{equation}
        \tau_i= A_i /R_{4f,5d}^2,
\label{tau}
\end{equation}
where $i$ denote a particular energy level.
The parameter $A_i$ is sensitive to the mixing of states (correlations between 
valence electrons) while there are also many-body corrections to
$R_{4f,5d}$ due to correlations between valence and core electrons.
Since the ratio of experimental and calculated lifetimes is almost 
the same for all three $^8P$ states (see Table \ref{lifetime}), it
is natural to assume that the most of discrepancy comes from
many-body corrections to $R_{4f,5d}$. Note, that the ratio of the
experimental and calculated lifetimes is even more
stable in the work of \citet{Mas}. The ratio is 
$3.0 \pm 0.3$ and the corresponding lifetimes are presented in column four
of Table \ref{lifetime}. Calculations in this work were also done with
the Cowan code, however the mixing of states was more carefully considered.

In the HF approximation, $R_{4f,5d}=0.77 a_B$. Now we 
calculate a correction to this value due to core polarization by
the dipole electric field of the emitted photon.
We do this in the random-phase approximation (RPA) using the
time-dependent Hartree-Fock method (TDHF) \cite{Dzu}. The TDHF equations can
be written in a form
\begin{equation}
        (\hat H_0 - \epsilon_i)\delta \psi_i = -(\hat f + \hat \delta V)\psi_i,
\label{RPA}
\end{equation}
where $\hat H_0$ is HF Hamiltonian. The single-electron orbital $\psi_i$
satisfies HF equation
\[      (\hat H_0 - \epsilon_i)\psi_i=0, \]
$\hat f$ is the operator of the external electric field,
$\delta \psi_i$ is a correction to the orbital $\psi_i$ due to external field
$\hat f$, and $\hat \delta V$ is the modification of the HF potential 
induced by corrections to the core states.
Equations (\ref{RPA}) are solved self-consistently
for all core states. Note that since Eu~III and Gd~IV are open-shell
systems, the same weighting procedure described in the previous 
section
must be applied to left-hand-side and right-hand-side of Eq.~(\ref{RPA}).
The transition amplitude between states $4f$ and
$5d$ in the RPA is
\begin{equation}
        \langle 4f|\hat f + \hat \delta V| 5d \rangle
\end{equation}
(the HF approximation corresponds to $\hat \delta V =0$).
Core polarization reduces the value of the  $R_{4f,5d}$ radial integral
bringing lifetimes into better agreement with experiment (see column ``RPA''
in Table~\ref{lifetime}). 

The remaining discrepancy should be attributed to
correlations. A detailed investigation of correlations leads beyond the
scope of the present work, but one should note that correlations increase the
density  of the external electron at short distances. Therefore, 
owing to normalization, it must decrease the density at large distances, 
thereby decreasing the value of the radial integral. 
Calculated lifetimes are also sensitive to mixing of states. 
Analysis of the RCI and Cowan code calculations shows that the smallest
mixing is for the $^8P_{9/2}$ state. Therefore, we can use this state 
to extract the value of $R_{4f,5d}$ that ensures the best fit of
the experimental data. This value is $R_{4f,5d}=0.41a_B$ for Eu$^{+2}$.
To calculate lifetimes which correspond to the ``best fit'' value of
$R_{4f,5d}$, one need only multiply the results from the Cowan code by a
factor of 3.6.
In summary, the values of $R_{4f,5d}$ for Eu~III and Gd~IV are:
\begin{center}
\begin{tabular}{ccc}\label{RGD}
         & Eu~III & Gd~IV \\
\hline
HF       &0.77$\, a_{B}$  &0.63$\, a_{B}$\\
RPA      &0.56$\, a_{B}$  &0.42$\, a_{B}$\\
Best fit &0.41$\, a_{B}$  &0.34$\, a_{B}$\\
\end{tabular}
\end{center}
\begin{table}
\caption{Lifetimes of $4f^65d~^8P_{5/2,7/2,9/2}$ states of Eu~III (ns).}
\label{lifetime}
\begin{ruledtabular}
\begin{tabular}{ccccc}
            & Exp.\footnotemark[1] & Cowan\footnotemark[2] &
Cowan\footnotemark[3] & RPA\footnotemark[4] \\
\hline
$^8P_{5/2}$ & 65(7) & 16 & 24 & 30  \\
$^8P_{7/2}$ & 46(5) & 15 & 15 & 28  \\
$^8P_{9/2}$ & 36(4) & 10 & 11 & 19  \\
\end{tabular}
\end{ruledtabular}
\footnotetext[1]{\citet{tau}}
\footnotetext[2]{this work, $R_{4f,5d}=0.77a_B$}
\footnotetext[3]{\citet{Mas}}
\footnotetext[4]{this work, $R_{4f,5d}=0.56a_B$}
\end{table}

\section{results}
\label{results}

 In Table~\ref{tab-eu}, we list and compare energies of
$4f^7$ and $4f^65d$ states in Eu~III calculated using the RCI code
and the Cowan code. Energies are given relative to the ground state
$4f^7\ ^8S_{7/2}$.
As mentioned above, both codes permit us to
obtain results that are generally in good agreement with
experimental energies by scaling the electrostatic Slater
parameters to simulate correlation effects
(Refs.~\cite{pindzola,sataka}).  We use the scaling factor of 0.8
in RCI code and 0.85 in the Cowan code. Also the  energies of the
$4f^65d\ LSJ$ levels are shifted by 13500 cm$^{-1}$ in the Cowan code
relative to the ground state $4f^7\ ^8S_{7/2}$. In the RCI code
we don't shift the energies but modify the $5d$ state as was described in
Section \ref{energies}, to improve the energy interval between the 
$4f^7$ and $4f^65d$ configurations.
In Table~\ref{tab-eu},  energies of the $4f^7\ LSJ$,
$4f^65d\ LSJ$, and $4f^66s\ LSJ$  levels in Eu~III are compared with
recommended data from the National Institute for Standards and Technology 
(NIST) by \citet{martin}. The 105 levels obtained
from spectral analysis  by \citet{eu}
given in the NIST publication, classify about 300 of the observed
lines.  It should be noted that the spectral analysis in
\cite{eu} was based on the Cowan code, probably, a simpler
version than we use here.

It should be noted that we use different coupling schemes in RCI ($jj$
coupling) and Cowan code ($LS$ coupling) to build energy matrices.
We use, for convenience, $LS$ coupling labeling of states;
however, neither $jj$ nor $LS$ coupling can describe the physical states
properly. To combine together our results obtained with different
coupling schemes, we calculated Land\'{e} $g$-factors  for each level.
For low-lying states, the $g$-factors are very close to their non-relativistic
values and identification of levels is easy. However, higher in
the spectrum, strong mixing between states makes level identification
difficult. We restricted our calculations to levels which
are reliably identified in both calculations.

As can be seen from Table~\ref{tab-eu} that  results of both
calculations for Eu~III are in good agreement with one another and with
experiment.
This gives us confidence in similar calculations for Gd~IV.

The RCI calculations for Gd~IV are done in exactly the same way as
for Eu~III. All fitting parameters were chosen for Eu~III and
no {\it a priori} data on Gd~IV was used in the calculations.
In a sense, we can say that the Gd~IV calculations are predictive. 
They produce an energy spectrum of Gd~IV regardless of what is known about it.
In contrast, the calculations with the Cowan code are not exactly the
same for both ions. While we use the same scaling factor for Coulomb
integrals (0.85) the energy shift for the $4f^65d$ configuration is
larger for Gd~IV (18000 cm$^{-1}$) than for Eu (13500 cm$^{-1}$).
A larger energy shift is needed  to obtain good agreement with available
experimental data.

In Table~\ref{tab-gd}, we  compare 
energies of the $4f^7\ LSJ$ and $4f^65d\ LSJ$ levels with
available experimental data and and predicted data given by
\citet{Gde}. It can be seen from
 Table~\ref{tab-gd} that for the $4f^7$ configuration the energies
 obtained by Cowan code $E^{\rm C}$  are in
better agreement with energies from \cite{Gde} than
are energies  obtained by RCI code $E^{\rm M}$.
However, for the  $4f^65d$ configuration, results of both calculations
are in very good agreement with each other and with
\cite{Gde}.

In  Tables (\ref{tab-eu} and \ref{tab-gd}), we  present
lifetimes of the $4f^65d\ LSJ$ levels calculated using the Cowan code
with the HF value of the $R_{4f,5d}$ radial integral. 
To get more accurate predictions
for the lifetimes one should multiply the values presented in tables 
by the factor of 3.6 (see section~\ref{lifetime}).

\section{conclusion}
In a recent work on calculation of the EDM enhancement factor ($K$) in Gd~IV
 \cite{Buh}, the result was presented in a form of two different numbers:
$K_A \approx -6.4$ and $K_B \approx -3.3$. These two numbers were based on
different assumptions about the energy splitting between $4f$ and $5d$ states
of Gd~IV.
The first number, ($K_A$) corresponds to
$E_{5d} - E_{4f} \approx 40,000$ cm$^{-1}$ which is a result of extrapolation
from Eu~III. The second number ($K_B$) corresponds to
$E_{5d} - E_{4f} \approx 100,000$ cm$^{-1}$ which is based on available
experimental data for Gd~IV (too incomplete at that time to be fully trusted).

The present work clearly indicates that the correct energy
splitting is closer to $100,000$ {cm}$^{-1}$ and consequently,
the enhancement factor is rather -3.3.

Furthermore, an analysis of lifetimes of Eu~III suggests that 
core polarization by the electric field of an external photon is 
an important effect for both ions, Eu~III and Gd~IV. It
reduces the value of the  $R_{4f,5d}$ radial integral by a factor
of about 1.5. This effect was not included in the calculation of 
the EDM enhancement factor ($K_{EDM}$) \cite{Buh}. Only contributions  
proportional to the  $R_{4f,5d}$ radial integral were considered 
in that work. To include core polarization by the electric field, 
one should  divide the final answer of Ref.~\cite{Buh} by the factor of 1.5.
This leaves us with $K_{\rm EDM}=-2.2 (-3.3/1.5)$.


\begin{table*}
\caption{Energies  (cm$^{-1}$) and lifetimes $\tau$ (sec) for in
Eu III calculated calculated by
RCI ($E^{\rm M}$) and Cowan code ($E^{\rm C}$) in [$4f^7$ + $4f^66p$],
[$4f^65d$ + $4f^66s$ model space. Energies are given relative to
the ground states $4f^7\ ^8S_{7/2}$. Comparison with recommended
NIST data ($E^{\rm N}$) \protect\cite{martin}.
\label{tab-eu}}
\begin{ruledtabular}
\begin{tabular}{lrrrlrrrrlrrrr}
\multicolumn{1}{c}{} &
\multicolumn{3}{c}{cm$^{-1}$} &
\multicolumn{1}{c}{} &
\multicolumn{3}{c}{cm$^{-1}$} &
\multicolumn{1}{c}{sec} &
\multicolumn{1}{c}{} &
\multicolumn{3}{c}{cm$^{-1}$} &
\multicolumn{1}{c}{sec} \\
\multicolumn{1}{c}{$LSJ$} &
\multicolumn{1}{c}{$E^{\rm M}$} &
\multicolumn{1}{c}{$E^{\rm C}$} &
\multicolumn{1}{c}{$E^{\rm N}$} &
\multicolumn{1}{c}{$LSJ$} &
\multicolumn{1}{c}{$E^{\rm M}$} &
\multicolumn{1}{c}{$E^{\rm C}$} &
\multicolumn{1}{c}{$E^{\rm N}$} &
\multicolumn{1}{c}{$\tau^{\rm C}$}&
\multicolumn{1}{c}{$LSJ$} &
\multicolumn{1}{c}{$E^{\rm M}$} &
\multicolumn{1}{c}{$E^{\rm C}$} &
\multicolumn{1}{c}{$E^{\rm N}$} &
\multicolumn{1}{c}{$\tau^{\rm C}$} \\
\hline
\multicolumn{4}{c}{$4f^7$ states} &
\multicolumn{5}{c}{$4f^65d$ states} &
\multicolumn{5}{c}{$4f^65d$ states} \\
$^6P_{3/2} $& 34182& 30406&        & $^8H_{3/2} $&  34103&  33642&   33856& 1.503[ 1]&$^8P_{5/2} $&  48111&  41756&   39769& 1.638[-8]\\
$^6P_{5/2} $& 33691& 30001&  28629 & $^8H_{5/2} $&  34800&  34160&   34394& 1.221[ 0]&$^8P_{7/2} $&  49410&  42139&   40871& 1.485[-8]\\
$^6P_{7/2} $& 33319& 29581&  28200 & $^8H_{7/2} $&  35722&  34848&   35109& 4.196[-1]&$^8P_{9/2} $&  50613&  43423&   42084& 1.024[-8]\\
	    &      &      &        & $^8H_{9/2} $&  36833&  35680&   35972& 1.418[-1]&	          &	  &	  &	   &  	      \\
$^6I_{7/2} $& 34702& 32295&  31746 & $^8H_{11/2}$&  38101&  36634&   36962& 4.524[-2]&$^6P_{3/2} $&  43323&  39747&	   & 6.317[-6]\\
$^6I_{9/2} $& 34972& 32560&  31954 & $^8H_{13/2}$&  39512&  37696&   38067& 1.472[-2]&$^6P_{5/2} $&  45407&  40763&   40898& 3.977[-8]\\
$^6I_{11/2}$& 35217& 32783&  32180 & $^8H_{15/2}$&  41065&  38865&   38290& 5.781[-3]&$^6P_{7/2} $&  48088&  43250&   42530& 4.040[-8]\\
$^6I_{13/2}$& 35396& 32921&  32314 & $^8H_{17/2}$&  42784&  40158&   40659& 4.292[-3]&  	 &	  &	  &	   &	      \\
$^6I_{15/2}$& 35465& 32926&  32308 & 	        &    	 &       &        &          &$^6H_{5/2} $&  48728&  44055&   43396& 1.956[-6]\\
$^6I_{17/2}$& 35352& 32716&  32073 & $^8D_{3/2} $&  37444&  35902&   35627& 9.567[-5]&$^6H_{7/2} $&  49463&  44593&   43885& 1.837[-6]\\
	   &       &      &        & $^8D_{5/2} $&  38726&  37059&        & 1.455[-5]&$^6H_{9/2} $&  50336&  45240&   44554& 1.671[-6]\\
$^6D_{1/2} $& 39714& 37159&        & $^8D_{7/2} $&  40016&  38129&   38229& 4.369[-6]&$^6H_{11/2}$&  51311&  45975&   45313& 1.483[-6]\\
$^6D_{3/2} $& 40048& 37342&        & $^8D_{9/2} $&  41275&  39101&   39226& 3.804[-6]&$^6H_{13/2}$&  52362&  46786&   46150& 1.271[-6]\\
$^6D_{5/2} $& 40295& 37457&        & $^8D_{11/2}$&  42482&  40017&   40133& 3.052[-1]&$^6H_{15/2}$&  53495&  47678&   47069& 1.051[-6]\\
$^6D_{7/2} $& 40145& 37275&        & 	         &       &       &        &          &  	  &	  &	  &	   &	      \\
$^6D_{9/2} $& 39231& 36562&        & $^8F_{1/2} $&  40362&  38832&        & 2.978[-3]&$^6F_{1/2} $&  52135&  48094&	   & 4.873[-6]\\
	   &       &      &        & $^8F_{3/2} $&  40999&  39334&   39014& 5.622[-4]&$^6F_{3/2} $&  54591&  48700&	   & 3.291[-6]\\
$^6G_{3/2} $& 49215& 47912&        & $^8F_{5/2} $&  41813&  39968&   39636& 7.318[-6]&$^6F_{5/2} $&  55764&  47434&   46108& 1.865[-6]\\
$^6G_{5/2} $& 48780& 47186&        & $^8F_{7/2} $&  42741&  40692&   40372& 1.259[-5]&$^6F_{7/2} $&  56851&  48188&   46793& 1.495[-6]\\
$^6G_{7/2} $& 48539& 46501&        & $^8F_{9/2} $&  43751&  41474&   41150& 3.811[-5]&$^6F_{9/2} $&  57693&  49130&   47714& 9.094[-7]\\
$^6G_{9/2} $& 48578& 47555&        & $^8F_{11/2}$&  44824&  42294&   41988& 9.703[-4]&$^6F_{11/2}$&  59967&  50735&   49086& 7.764[-6]\\
$^6G_{11/2}$& 48743& 47645&        & $^8F_{13/2}$&  45957&  43138&   42850& 4.093[-4]&  	  &	  &	  &	   &	      \\
$^6G_{13/2}$& 49859& 49110&        & 	        &    	 &       &        &          &$^6D_{1/2} $&  64852&  46902&	   & 1.796[-6]\\
	   &       &      &        & $^8G_{1/2} $&  39692&  38440&   38050& 1.586[-3]&$^6D_{3/2} $&  65882&  46948&	   & 2.090[-6]\\
$^6F_{1/2} $& 52339& 49089&        & $^8G_{3/2} $&  39989&  38657&   38337& 9.284[-4]&$^6D_{5/2} $&  65929&  49445&   48496& 2.595[-6]\\
$^6F_{3/2} $& 52865& 51589&        & $^8G_{5/2} $&  40642&  39160&   38829& 1.417[-5]&$^6D_{7/2} $&  66715&  50206&   49293& 1.911[-6]\\
$^6F_{5/2} $& 53356& 51293&        & $^8G_{7/2} $&  41583&  39898&   39580& 1.581[-5]&$^6D_{9/2} $&  67322&  50846&   49957& 1.094[-6]\\  
$^6F_{7/2} $& 53699& 51605&        & $^8G_{9/2} $&  42747&  40814&   40518& 4.568[-5]&  	  &	  &	  &	   &	      \\
$^6F_{9/2} $& 53744& 51812&        & $^8G_{11/2}$&  44073&  41848&   41572& 3.865[-4]&$^6G_{3/2} $&  52052&  50504&	   & 8.288[-5]\\
$^6F_{11/2}$& 53341& 51165&        & $^8G_{13/2}$&  45474&  42926&   42658& 1.222[-4]&$^6G_{5/2} $&  52902&  51193&   49906& 2.662[-5]\\
	   &       &      &        & $^8G_{15/2}$&  46832&  43952&   43658& 3.452[-5]&$^6G_{7/2} $&  54046&  51896&	   & 1.733[-5]\\
           &       &      &        &   	         &       &       &        &   	     &$^6G_{9/2} $&  55329&  52533&	   & 1.639[-5]\\
           &       &      &        &   	         &       &       &        &          &$^6G_{11/2}$&  57039&  53055&   51651& 4.744[-5]\\
           &       &      &        &   	         &       &       &        &   	     &$^6G_{13/2}$&  60645&  53464&   52100& 1.108[-4]\\
\end{tabular}
\end{ruledtabular}
\end{table*}
\newpage

\begin{table*}
 \caption{Energies
(cm$^{-1}$) and lifetimes $\tau$ (sec) for in Gd IV calculated by
RCI ($E^{\rm M}$) and Cowan code ($E^{\rm C}$). Energies are
given relative to the ground states $4f^7\ ^8S_{7/2}$. Comparison
with experimental data from Ref.~\protect\cite{Gde}
($E^{\rm exp}$).
 \label{tab-gd}}
\begin{ruledtabular}
\begin{tabular}{lrrrlrrrlrrrr}
\multicolumn{1}{c}{} &
\multicolumn{3}{c}{cm$^{-1}$} &
\multicolumn{1}{c}{} &
\multicolumn{2}{c}{cm$^{-1}$} &
\multicolumn{1}{c}{sec} &
\multicolumn{1}{c}{} &
\multicolumn{3}{c}{cm$^{-1}$} &
\multicolumn{1}{c}{sec} \\
\multicolumn{1}{c}{$LSJ$} &
\multicolumn{1}{c}{$E^{\rm M}$} &
\multicolumn{1}{c}{$E^{\rm C}$} &
\multicolumn{1}{c}{$E^{\rm exp}$} &
\multicolumn{1}{c}{$LSJ$} &
\multicolumn{1}{c}{$E^{\rm M}$} &
\multicolumn{1}{c}{$E^{\rm C}$} &
\multicolumn{1}{c}{$\tau^{\rm C}$}&
\multicolumn{1}{c}{$LSJ$} &
\multicolumn{1}{c}{$E^{\rm M}$} &
\multicolumn{1}{c}{$E^{\rm C}$} &
\multicolumn{1}{c}{$E^{\rm exp}$} &
\multicolumn{1}{c}{$\tau^{\rm C}$}\\
\hline
\multicolumn{4}{c}{$4f^7$ states} &
\multicolumn{4}{c}{$4f^65d$ states} &
\multicolumn{5}{c}{$4f^65d$ states} \\
$^6P_{3/2} $&  38308  & 34114 &  33262&$^8H_{3/2} $&   92479&   98073&  3.364[-5] &$^6P_{3/2} $&   103091& 105613&  104264 &2.131[-8]\\
$^6P_{5/2} $&  37638  & 33577 &  32680&$^8H_{5/2} $&   93338&   98750&  1.624[-5] &$^8P_{5/2} $&   105108& 106266&  106493 &1.696[-9]\\
$^6P_{7/2} $&  37103  & 33018 &  32084&$^8H_{7/2} $&   94467&   99641&  9.024[-6] &$^6P_{7/2} $&   107493& 109547&  109005 &4.311[-9]\\
            &         &       &       &$^8H_{9/2} $&   95821&  100711&  5.447[-6] &            &         &       &         &         \\
$^6I_{7/2} $&  38833  & 36109 &  35808&$^8H_{11/2}$&   97368&  101933&  3.541[-6] &$^6H_{5/2} $&   108974& 110344&         &1.093[-8]\\
$^6I_{9/2} $&  39191  & 36468 &  36151&$^8H_{13/2}$&   99097&  103295&  2.527[-6] &$^6H_{7/2} $&   109790& 110989&         &1.086[-8]\\
$^6I_{11/2}$&  39504  & 36766 &  36430&$^8H_{15/2}$&  101027&  104809&  2.160[-6] &$^6H_{9/2} $&   110760& 111763&         &1.075[-8]\\
$^6I_{13/2}$&  39722  & 36950 &  36508&$^8H_{17/2}$&  103239&  106541&  6.287[-6] &$^6H_{11/2}$&   111838& 112641&         &1.067[-8]\\
$^6I_{15/2}$&  39782  & 36957 &  36547&            &        &        &            &$^6H_{13/2}$&   113003& 113620&         &1.069[-8]\\
$^6I_{17/2}$&  39586  & 36677 &  36206&$^8D_{3/2} $&   96222&  100611&  6.089[-8] &$^6H_{15/2}$&   114270& 114729&         &1.089[-8]\\
            &         &       &       &$^8D_{5/2} $&   98072&  102237&  1.195[-7] &            &         &       &         &         \\
$^6D_{1/2} $&  44618  & 41738 &  40444&$^8D_{7/2} $&   99762&  103840&  4.803[-7] &$^6F_{1/2} $&   112718& 115165&         &1.405[-8]\\
$^6D_{3/2} $&  45060  & 41983 &  40694&$^8D_{9/2} $&  101204&  105041&  4.429[-7] &$^6F_{3/2} $&   112551& 115883&         &1.317[-8]\\
$^6D_{5/2} $&  45363  & 42131 &  40857&$^8D_{11/2}$&  102467&  106068&  2.438[-6] &$^6F_{5/2} $&   113476& 114242&  111745 &1.068[-8]\\
$^6D_{7/2} $&  45120  & 41877 &  40599&            &        &        &            &$^6F_{7/2} $&   114715& 115140&  113129 &9.812[-9]\\
$^6D_{9/2} $&  43876  & 40934 &  39508&$^8F_{1/2} $&   99582&  103800&  7.829[-7] &$^6F_{9/2} $&   116120& 116239&  114214 &9.076[-9]\\
            &         &       &       &$^8F_{3/2} $&  100315&  104908&  8.558[-7] &$^6F_{11/2}$&   118102& 118166&         &1.166[-8]\\
$^6G_{3/2} $&  54999  & 53498 &  50633&$^8F_{5/2} $&  101260&  104774&  1.958[-7] &            &         &       &         &         \\
$^6G_{5/2} $&  54398  & 52556 &  49825&$^8F_{7/2} $&  102350&  106595&  1.884[-7] &$^6D_{1/2} $&   114452& 113648&         &1.207[-8]\\
$^6G_{7/2} $&  54061  & 51626 &  49526&$^8F_{9/2} $&  103560&  107573&  7.487[-7] &$^6D_{3/2} $&   115542& 113646&         &1.196[-8]\\
$^6G_{9/2} $&  54132  & 53018 &  49652&$^8F_{11/2}$&  104888&  108599&  2.167[-6] &$^6D_{5/2} $&   116881& 116788&         &1.229[-8]\\
$^6G_{11/2}$&  54277  & 53113 &  49652&$^8F_{13/2}$&  106352&  109718&  4.888[-7] &$^6D_{7/2} $&   118099& 117691& 116230  &1.156[-8]\\
$^6G_{13/2}$&  55684  & 54908 &  51360&            &        &        &            &$^6D_{9/2} $&   119010& 118415& 117229  &1.059[-8]\\
            &         &       &       &$^8G_{1/2} $&   98870&  104285&  1.263[-7] &            &         &       &         &         \\
$^6F_{1/2} $&  58661  & 55022 &       &$^8G_{3/2} $&   99263&  104094&  6.047[-7] &$^6G_{3/2} $&   116779& 117952&         &1.700[-8]\\
$^6F_{3/2} $&  59353  & 58025 &       &$^8G_{5/2} $&  100104&  105690&  8.230[-8] &$^6G_{5/2} $&   118903& 118862&  118109 &1.553[-8]\\
$^6F_{5/2} $&  59953  & 57566 &       &$^8G_{7/2} $&  101315&  105781&  2.170[-7] &$^6G_{7/2} $&   119962& 119714&  119292 &1.528[-8]\\
$^6F_{7/2} $&  60303  & 57909 &       &$^8G_{9/2} $&  102816&  107028&  5.390[-7] &$^6G_{9/2} $&   120838& 120449&  120220 &1.513[-8]\\
$^6F_{9/2} $&  60310  & 58200 &       &$^8G_{11/2}$&  104455&  108380&  5.352[-7] &$^6G_{11/2}$&   121510& 121003&  121063 &1.496[-8]\\
$^6F_{11/2}$&  59774  & 57330 &       &$^8G_{13/2}$&  106038&  109641&  2.946[-7] &$^6G_{13/2}$&   122130& 121368&  121725 &1.393[-8]\\
            &         &       &       &$^8G_{15/2}$&  107581&  110766&  8.983[-8] &            &         &       &         &         \\
$^4N_{17/2}$&  60545  &  55382&       &            &        &        &            &$^6G_{3/2} $&   118067& 119462&         &7.208[-8]\\
$^4N_{19/2}$&  61512  &  56379&       &$^8P_{5/2} $&  107010&  107583&  2.264[-9] &$^6G_{5/2} $&   119183& 121201&         &1.101[-7]\\
$^4N_{21/2}$&  62009  &  56524&       &$^8P_{7/2} $&  108921&  107524&  1.180[-9] &$^6G_{7/2} $&   122136& 123606&         &1.262[-7]\\
$^4N_{23/2}$&  61817  &  56827&       &$^8P_{9/2} $&  109318&  108884&  9.829[-10]&$^6G_{9/2} $&   125622& 126445&         &1.554[-7]\\
            &         &       &       &            &        &        &            &$^6G_{11/2}$&   128024& 129607&         &1.376[-7]\\
            &         &       &       &            &        &        &            &$^6G_{13/2}$&   129323&       &         &         \\
\end{tabular}
\end{ruledtabular}
\end{table*}

\begin{acknowledgments}
One of the authors (V.D.) is grateful to the
Physics Department of the University of
Notre Dame for the hospitality and support during
his visit in May, 2002.
The work of W.R.J. was supported in part by National
Science Foundation Grant No.\ PHY-01-39928. U.I.S. acknowledges
partial support by Grant No.\ B516165 from Lawrence Livermore
National Laboratory. 
\end{acknowledgments}


\begin{thebibliography}{21}
\expandafter\ifx\csname natexlab\endcsname\relax\def\natexlab#1{#1}\fi
\expandafter\ifx\csname bibnamefont\endcsname\relax
  \def\bibnamefont#1{#1}\fi
\expandafter\ifx\csname bibfnamefont\endcsname\relax
  \def\bibfnamefont#1{#1}\fi
\expandafter\ifx\csname citenamefont\endcsname\relax
  \def\citenamefont#1{#1}\fi
\expandafter\ifx\csname url\endcsname\relax
  \def\url#1{\texttt{#1}}\fi
\expandafter\ifx\csname urlprefix\endcsname\relax\def\urlprefix{URL }\fi
\providecommand{\bibinfo}[2]{#2}
\providecommand{\eprint}[2][]{\url{#2}}

\bibitem[{\citenamefont{Lamoreaux}()}]{Lam}
\bibinfo{author}{\bibfnamefont{S.~K.} \bibnamefont{Lamoreaux}},
  \eprint{nucl-ex/0109014}.

\bibitem[{\citenamefont{Hunter}()}]{Hun}
\bibinfo{author}{\bibfnamefont{L.~R.} \bibnamefont{Hunter}},
  \bibinfo{note}{workshop on {\it Tests of Fundamental Symmetries in Atoms and
  Molecules}, Harvard, (2001) available online
  http://itamp.harvard.edu/fundamentalworkshop.html}.

\bibitem[{\citenamefont{Khriplovich and Lamoreaux}(1997)}]{KL}
\bibinfo{author}{\bibfnamefont{I.~B.} \bibnamefont{Khriplovich}}
  \bibnamefont{and} \bibinfo{author}{\bibfnamefont{S.~K.}
  \bibnamefont{Lamoreaux}}, \emph{\bibinfo{title}{CP Violation Without
  Strangeness}} (\bibinfo{publisher}{Springer}, \bibinfo{address}{Berlin},
  \bibinfo{year}{1997}).

\bibitem[{\citenamefont{Regan et~al.}(2002)\citenamefont{Regan, Commins,
  Schmidt, and DeMille}}]{Com}
\bibinfo{author}{\bibfnamefont{B.~C.} \bibnamefont{Regan}},
  \bibinfo{author}{\bibfnamefont{E.~D.} \bibnamefont{Commins}},
  \bibinfo{author}{\bibfnamefont{C.~J.} \bibnamefont{Schmidt}},
  \bibnamefont{and} \bibinfo{author}{\bibfnamefont{D.}~\bibnamefont{DeMille}},
  \bibinfo{journal}{Phys. Rev. Lett.} \textbf{\bibinfo{volume}{88}},
  \bibinfo{pages}{071805} (\bibinfo{year}{2002}).

\bibitem[{\citenamefont{DeMille}()}]{DeM}
\bibinfo{author}{\bibfnamefont{D.}~\bibnamefont{DeMille}},
  \bibinfo{note}{workshop on {\it Tests of Fundamental Symmetries in Atoms and
  Molecules}, Harvard, (2001) available online
  http://itamp.harvard.edu/fundamentalworkshop.html}.

\bibitem[{\citenamefont{Hudson et~al.}()\citenamefont{Hudson, Sauer, Tarbutt,
  and Hinds}}]{Hinds}
\bibinfo{author}{\bibfnamefont{J.~J.} \bibnamefont{Hudson}},
  \bibinfo{author}{\bibfnamefont{B.~E.} \bibnamefont{Sauer}},
  \bibinfo{author}{\bibfnamefont{M.~R.} \bibnamefont{Tarbutt}},
  \bibnamefont{and} \bibinfo{author}{\bibfnamefont{E.~A.} \bibnamefont{Hinds}},
  \eprint{hep-ex/0202014}.

\bibitem[{\citenamefont{Shapiro}(1968)}]{Sh}
\bibinfo{author}{\bibfnamefont{F.~L.} \bibnamefont{Shapiro}},
  \bibinfo{journal}{Sov.\ Phys.\ Usp} \textbf{\bibinfo{volume}{11}},
  \bibinfo{pages}{345} (\bibinfo{year}{1968}).

\bibitem[{\citenamefont{Vasil'ev and Kolycheva}(1978)}]{VK}
\bibinfo{author}{\bibfnamefont{B.~V.} \bibnamefont{Vasil'ev}} \bibnamefont{and}
  \bibinfo{author}{\bibfnamefont{E.~V.} \bibnamefont{Kolycheva}},
  \bibinfo{journal}{Sov.\ Phys.\ JETP} \textbf{\bibinfo{volume}{47}},
  \bibinfo{pages}{243} (\bibinfo{year}{1978}).

\bibitem[{\citenamefont{Buhmann et~al.}()\citenamefont{Buhmann, Dzuba, and
  Sushkov}}]{Buh}
\bibinfo{author}{\bibfnamefont{S.~Y.} \bibnamefont{Buhmann}},
  \bibinfo{author}{\bibfnamefont{V.~A.} \bibnamefont{Dzuba}}, \bibnamefont{and}
  \bibinfo{author}{\bibfnamefont{O.~P.} \bibnamefont{Sushkov}},
  \eprint{physics/0204076}.

\bibitem[{\citenamefont{Kuenzi et~al.}()\citenamefont{Kuenzi, Sushkov, Dzuba,
  and Cadogan}}]{Kuen}
\bibinfo{author}{\bibfnamefont{S.~A.} \bibnamefont{Kuenzi}},
  \bibinfo{author}{\bibfnamefont{O.~P.} \bibnamefont{Sushkov}},
  \bibinfo{author}{\bibfnamefont{V.~A.} \bibnamefont{Dzuba}}, \bibnamefont{and}
  \bibinfo{author}{\bibfnamefont{J.~M.} \bibnamefont{Cadogan}},
  \eprint{cond-mat/0205113}.

\bibitem[{\citenamefont{Paoletti}(1978)}]{GGG}
\bibinfo{author}{\bibfnamefont{A.}~\bibnamefont{Paoletti}},
  \emph{\bibinfo{title}{Physics of Magnetic Garnet}}
  (\bibinfo{publisher}{North-Holland}, \bibinfo{address}{Amsterdam},
  \bibinfo{year}{1978}).

\bibitem[{\citenamefont{Kielkopf and Crosswhite}(1970)}]{Gde}
\bibinfo{author}{\bibfnamefont{J.~F.} \bibnamefont{Kielkopf}} \bibnamefont{and}
  \bibinfo{author}{\bibfnamefont{H.~M.} \bibnamefont{Crosswhite}},
  \bibinfo{journal}{J. Opt.\ Soc.\ Am.} \textbf{\bibinfo{volume}{60}},
  \bibinfo{pages}{347} (\bibinfo{year}{1970}).

\bibitem[{\citenamefont{Zhiguo et~al.}(2000)\citenamefont{Zhiguo, Li, Lundberg,
  Zhang, Dai, Zhankui, and Svanberg}}]{tau}
\bibinfo{author}{\bibfnamefont{Z.}~\bibnamefont{Zhiguo}},
  \bibinfo{author}{\bibfnamefont{Z.~S.} \bibnamefont{Li}},
  \bibinfo{author}{\bibfnamefont{H.}~\bibnamefont{Lundberg}},
  \bibinfo{author}{\bibfnamefont{K.~Y.} \bibnamefont{Zhang}},
  \bibinfo{author}{\bibfnamefont{Z.~W.} \bibnamefont{Dai}},
  \bibinfo{author}{\bibfnamefont{J.}~\bibnamefont{Zhankui}}, \bibnamefont{and}
  \bibinfo{author}{\bibfnamefont{S.}~\bibnamefont{Svanberg}},
  \bibinfo{journal}{J. Phys.\ B} \textbf{\bibinfo{volume}{33}},
  \bibinfo{pages}{521} (\bibinfo{year}{2000}).

\bibitem[{\citenamefont{Cowan}(1981)}]{cowan}
\bibinfo{author}{\bibfnamefont{R.~D.} \bibnamefont{Cowan}},
  \emph{\bibinfo{title}{The Theory of Atomic Structure and Spectra}}
  (\bibinfo{publisher}{University of California Press},
  \bibinfo{address}{Berkeley}, \bibinfo{year}{1981}).

\bibitem[{\citenamefont{Dzuba et~al.}(1996)\citenamefont{Dzuba, Flambaum, and
  Kozlov}}]{kozlov}
\bibinfo{author}{\bibfnamefont{V.~A.} \bibnamefont{Dzuba}},
  \bibinfo{author}{\bibfnamefont{V.~V.} \bibnamefont{Flambaum}},
  \bibnamefont{and} \bibinfo{author}{\bibfnamefont{M.~G.}
  \bibnamefont{Kozlov}}, \bibinfo{journal}{Phys.\ Rev.\ A}
  \textbf{\bibinfo{volume}{54}}, \bibinfo{pages}{3948} (\bibinfo{year}{1996}).

\bibitem[{\citenamefont{Mashonkina et~al.}(2002)\citenamefont{Mashonkina,
  Ryabtsev, and Ryabchikova}}]{Mas}
\bibinfo{author}{\bibfnamefont{L.~I.} \bibnamefont{Mashonkina}},
  \bibinfo{author}{\bibfnamefont{A.~N.} \bibnamefont{Ryabtsev}},
  \bibnamefont{and} \bibinfo{author}{\bibfnamefont{T.~A.}
  \bibnamefont{Ryabchikova}}, \bibinfo{journal}{Astronomy Letters}
  \textbf{\bibinfo{volume}{28}}, \bibinfo{pages}{34} (\bibinfo{year}{2002}).

\bibitem[{\citenamefont{Dzuba et~al.}(1987)\citenamefont{Dzuba, Flambaum,
  Silvestrov, and Sushkov}}]{Dzu}
\bibinfo{author}{\bibfnamefont{V.~A.} \bibnamefont{Dzuba}},
  \bibinfo{author}{\bibfnamefont{V.~V.} \bibnamefont{Flambaum}},
  \bibinfo{author}{\bibfnamefont{P.~G.} \bibnamefont{Silvestrov}},
  \bibnamefont{and} \bibinfo{author}{\bibfnamefont{O.~P.}
  \bibnamefont{Sushkov}}, \bibinfo{journal}{J. Phys.\ B}
  \textbf{\bibinfo{volume}{20}}, \bibinfo{pages}{1399} (\bibinfo{year}{1987}).

\bibitem[{\citenamefont{Pindzola et~al.}(1994)\citenamefont{Pindzola, Gorczyca,
  Badnell, Griffin, Stenke, Hofmann, B.Weissbecker, Tinschert, Salzborn,
  M{\"{u}}ller et~al.}}]{pindzola}
\bibinfo{author}{\bibfnamefont{M.~S.} \bibnamefont{Pindzola}},
  \bibinfo{author}{\bibfnamefont{T.~W.} \bibnamefont{Gorczyca}},
  \bibinfo{author}{\bibfnamefont{N.~R.} \bibnamefont{Badnell}},
  \bibinfo{author}{\bibfnamefont{D.~C.} \bibnamefont{Griffin}},
  \bibinfo{author}{\bibfnamefont{M.}~\bibnamefont{Stenke}},
  \bibinfo{author}{\bibfnamefont{G.}~\bibnamefont{Hofmann}},
  \bibinfo{author}{\bibnamefont{B.Weissbecker}},
  \bibinfo{author}{\bibfnamefont{K.}~\bibnamefont{Tinschert}},
  \bibinfo{author}{\bibfnamefont{E.}~\bibnamefont{Salzborn}},
  \bibinfo{author}{\bibfnamefont{A.}~\bibnamefont{M{\"{u}}ller}},
  \bibnamefont{et~al.}, \bibinfo{journal}{Phys.\ Rev.\ A}
  \textbf{\bibinfo{volume}{49}}, \bibinfo{pages}{933} (\bibinfo{year}{1994}).

\bibitem[{\citenamefont{Sataka et~al.}(2002)\citenamefont{Sataka, Imai,
  Kawatsura, Komaki, Tawara, Vasilyev, and Safronova}}]{sataka}
\bibinfo{author}{\bibfnamefont{M.}~\bibnamefont{Sataka}},
  \bibinfo{author}{\bibfnamefont{M.}~\bibnamefont{Imai}},
  \bibinfo{author}{\bibfnamefont{K.}~\bibnamefont{Kawatsura}},
  \bibinfo{author}{\bibfnamefont{K.}~\bibnamefont{Komaki}},
  \bibinfo{author}{\bibfnamefont{H.}~\bibnamefont{Tawara}},
  \bibinfo{author}{\bibfnamefont{A.}~\bibnamefont{Vasilyev}}, \bibnamefont{and}
  \bibinfo{author}{\bibfnamefont{U.~I.} \bibnamefont{Safronova}},
  \bibinfo{journal}{Phys.\ Rev.\ A} \textbf{\bibinfo{volume}{65}},
  \bibinfo{pages}{052704} (\bibinfo{year}{2002}).

\bibitem[{\citenamefont{Martin et~al.}(1978)\citenamefont{Martin, Zalubas, and
  Hagan}}]{martin}
\bibinfo{author}{\bibfnamefont{W.~C.} \bibnamefont{Martin}},
  \bibinfo{author}{\bibfnamefont{R.}~\bibnamefont{Zalubas}}, \bibnamefont{and}
  \bibinfo{author}{\bibfnamefont{L.}~\bibnamefont{Hagan}},
  \emph{\bibinfo{title}{Atomic Energy Levels - The Rare-Earth Elements}}
  (\bibinfo{publisher}{U. S. Government Printing Office},
  \bibinfo{address}{Washington DC}, \bibinfo{year}{1978}).

\bibitem[{\citenamefont{Sugar and Spector}(1974)}]{eu}
\bibinfo{author}{\bibfnamefont{J.}~\bibnamefont{Sugar}} \bibnamefont{and}
  \bibinfo{author}{\bibfnamefont{N.}~\bibnamefont{Spector}},
  \bibinfo{journal}{J. Opt.\ Soc.\ Am} \textbf{\bibinfo{volume}{64}},
  \bibinfo{pages}{1484} (\bibinfo{year}{1974}).

\end{thebibliography}

\end{document}